\documentclass[a4paper,fleqn,usenatbib]{mnras}

\usepackage[T1]{fontenc}
\usepackage{ae,aecompl}

\usepackage{graphicx}	
\usepackage{amsmath}	
\usepackage{amssymb}	

\newcommand{\msun}{\hbox{M$_{\odot}$}}
\newcommand{\kms}{\hbox{km s$^{-1}$}}

\title[Is {\rm[Mg/Fe]} a good proxy for galaxy formation time-scales?]{Revisiting the classics: Is [Mg/Fe] a good proxy for galaxy formation time-scales?}

\author[I. Mart\'in-Navarro]{
Ignacio Mart\'in-Navarro$^{1,2}$\thanks{E-mail: imartin@iac.es}
\\
$^{1}$Instituto de Astrof\'isica de Canarias, E-38200 La Laguna, Tenerife, Spain\\
$^{2}$Departamento de Astrof\'isica, Universidad de La Laguna, E-38205 La Laguna, Tenerife, Spain\\
}

\date{Accepted XXX. Received YYY; in original form ZZZ}

\pubyear{2015}

\begin{document}
\label{firstpage}
\pagerange{\pageref{firstpage}--\pageref{lastpage}}
\maketitle

\begin{abstract}
In the local Universe, massive early-type galaxies exhibit enhanced [Mg/Fe] ratios, which has been traditionally interpreted as the result of a rapid ($\tau \lesssim 1$ Gyr) collapse. However, recent claims of a non-universal, steep initial mass function call for a revision of this standard interpretation. In the present work we show how the simultaneous consideration of a high [Mg/Fe] and a steep IMF slope would imply unreasonably short ($\tau \sim 7$ Myr) and intense (SFR $\sim 10^{5}$ \msun\,yr$^{-1}$) formation events for massive early-type galaxies. We discuss possible caveats and explanations to this apparent inconsistency, and we suggest that further IMF determinations, both in the local Universe and at high redshift, are necessary to better understand the problem.
\end{abstract}



\section{Introduction}

Early-type galaxies (ETGs) have been traditionally considered as landmarks for stellar population analysis since, unlike other morphological types, they are well represented by simple star formation histories. Observations suggest that ETGs harbour old and metal-rich populations, with more massive systems being older and more metallic than low-mass ETGs \citep{Trager00,thomas05}. The study of absorption lines in the spectrum of ETGs has also revealed a tight correlation between the strength of Mg features and galaxy velocity dispersion or luminosity \citep{Peletier, Bender, Worthey00}, which has been interpreted as an increasing $\alpha$-elements to iron ratio with increasing galaxy mass \citep{Worthey92,Thomas99}. 

To explain the over-abundant [Mg/Fe] ratio in massive ETGs, three main mechanisms were proposed \citep{Worthey92}: (1) different formation time-scales compared to the solar neighbourhood, (2) a non-universal stellar initial mass function (IMF), flatter than observed in the Milky Way, and (3) selective mass loss. Whereas the selective loss mechanism predicts an anti-correlation between [Mg/Fe] and galaxy velocity dispersion, a universal, Milky-Way like IMF was strongly supported by observations of nearby resolved systems  \citep{kroupa,bastian}. Thus, differences in the formation time-scales became the preferred explanation to the observed [Mg/Fe] enhancement in massive ETGs.

Based on these observational results, a broad picture of ETGs formation emerged: massive ETGs formed at high-redshift, in short and intense formation episodes and within massive halos. This would lead to old, [Mg/Fe] enhanced and metal-rich systems, as supported by observations. Moreover, since galaxy mass ultimately regulates all these processes, the age--, metallicity-- and [Mg/Fe]--velocity dispersion relations are naturally explained within this framework. All these ideas were synthesised by \citet{thomas05}, where, assuming a Milky-Way like IMF, they showed that not only the age and the metallicity, but also the formation-time scale can be retrieved from the optical spectrum of an ETG. 

Recently, a new piece of observational evidence has been added to the puzzle: the IMF of ETGs is not universal \citep{vandokkum}. Both strong- \citep{Treu} and micro-lensing \citep{paul}, dynamical modelling \citep{cappellari} and stellar population analysis \citep{ferreras} suggest a systematic departure from a Milky-Way like IMF when increasing galaxy mass. In this regard, a detailed analysis of IMF-sensitive features \citep{labarbera,Spiniello2013} indicates that massive ETGs ($\sigma \sim 300$ \kms) host an enhanced population of M-dwarf stars, transitioning to a Milky-Way like IMF for low-mass ($\sigma \sim 100$ \kms) objects.

Since the IMF determines, not only the number of massive stars, but also the amount of gas locked in form of low-mass stars, the assumption of a steep IMF would have strong implications on the chemical evolution of a galaxy. In this regard, it has been claimed \citep{Calura,weidner:13,Ferreras15} that such bottom-heavy IMF is in tension with the observed chemical properties of massive ETGs, as a metal enrichment regulated by a steep IMF is not able to produce the high metallicity measured in this type of objects. In addition, \citet{anna13} have shown that the star formation histories inferred from full-spectral fitting depends on the adopted IMF, as a reddening in the spectrum can be either caused by a high fraction of low-mass stars or actually by an old stellar population.

In this letter we will focus on the influence of a non-universal, steep IMF on the [Mg/Fe] ratio, traditionally considered as a reliable proxy for the formation time-scale. We will show that our current observational framework, where massive ETGs exhibit simultaneously bottom-heavy IMFs and enhanced [Mg/Fe] ratios, would lead to extremely short ($\tau \lesssim$ 0.01 Gyr) formation episodes, and consequently to unobserved star-formation-rates (SFR $\sim 10^{5}$ \msun\,yr$^{-1}$) at high-redshift.

\section{Data}
Without loss of generality, our analysis is compared to the benchmark study of \citet{thomas05}. They gathered a sample of 124 ETGs covering a wide range in stellar mass (100 \kms $\lesssim \sigma \lesssim $ 300 \kms), located in low- and high-density environments. Since it has no significant influence on the [Mg/Fe] ratio, throughout this paper we will not consider environmental effects. Notice, however, that ages of ETGs strongly depend on the environment (being older in denser regions). 

To reproduce and fairly compare the results of \citet{thomas05}, we adopted their chemical evolutionary models, first described in \citet{Thomas99}. These models explored the IMF, time-scale, [Mg/Fe] parameter space for two extreme cases: {\it a fast clumpy collapse and the merger of two spirals}. Moreover, they computed the chemical evolution on the basis of two stellar yields recipes, \citet{ww95} and \citet{tnh96}. Following \citet{thomas05}, we stick to the clumpy collapse plus \citet{tnh96} stellar yields models\footnote{In \S\ref{sec:disc} we discuss the importance of this model choice.}. 

Finally, we used \citet{labarbera} as our reference for the IMF--$\sigma$ relation. Both \citet{thomas05} and \citet{labarbera} cover a similar mass range, although the latter relies on fibre-based stacked spectra. To be consistent with \citet{Thomas99}, we adopted the \mbox{IMF--$\sigma$} relation resultant from a unimodal (single power-law like) IMF parametrization.

\section{Results and comparison}

\begin{figure}
\begin{center}
\includegraphics[width=8.5cm]{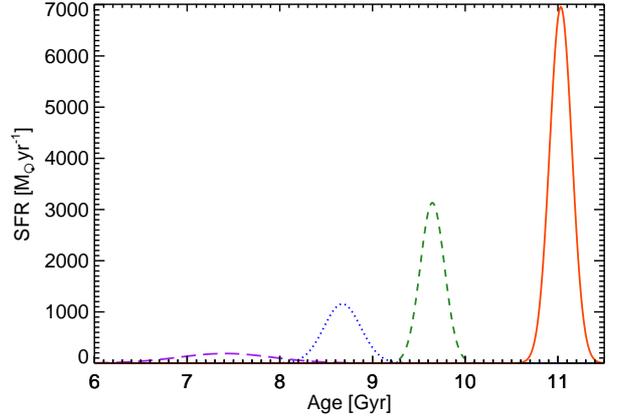}
\end{center}
\caption{Schematic representation of the classical ETGs formation scenario, showing the star formation histories (parametrized as gaussian functions) for different galaxies with different velocity dispersions (i.e., different masses). The red, green, blue and purple lines correspond to the typical star formation history of a galaxy with $\sigma \sim 300$\kms, $\sigma \sim 270$\kms, $\sigma \sim 230$\kms and $\sigma \sim 190$\kms, respectively. Gaussians are centred at the the mean age of each $\sigma$ bin, as inferred from the analysis of Balmer optical lines, whereas the width of the distributions is given by the [Mg/Fe] proxy, assuming a fixed Milky Way like IMF. In this standard picture, more massive ETGs formed earlier and faster than low-mass ETGs.}
\label{fig:thomas}
\end{figure}

Following \citet{thomas05}, Fig.~\ref{fig:thomas} schematically represents the current ETGs formation scenario. Coloured curves trace the typical star formation histories of ETGs with different masses. For simplicity, the star formation histories are shown as a gaussian function defined by:
\begin{equation}
 \mathrm{SFR(age)} = \frac{M_\star}{\tau} \cdot e^{{{ - \left( {\mathrm{age} - \langle\mathrm{age}\rangle_\sigma } \right)^2 } \mathord{\left/ {\vphantom {{ - \left( {x - \mu } \right)^2 } {2\tau ^2 }}} \right. \kern-\nulldelimiterspace} {2\tau ^2 }}}
\end{equation}
The mean corresponds to the mean age in that mass bin, $\langle\mathrm{age}\rangle_\sigma$. The standard deviation is given by the typical formation time-scale $\tau$, as derived from the [Mg/Fe] ratio assuming a Milky Way like IMF.

\begin{figure}
\begin{center}
\includegraphics[width=8.5cm]{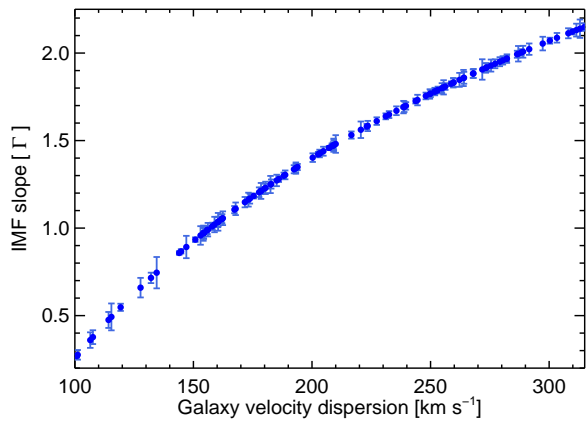}
\end{center}
\caption{Estimated IMF slope for the sample of \citet{thomas05}, following the IMF-$\sigma$ relation of \citet{labarbera}. Galaxies with a velocity dispersion $\sigma \sim 150$ \kms are well represented by a Milky Way like IMF, but more massive ETGs require steeper IMFs. We used these estimations of the IMF slope to recalculate the typical formation time-scale of each galaxy in the sample.}
\label{fig:francesco}
\end{figure}

To understand how does Fig.~\ref{fig:thomas} depend on the adopted IMF, we first used the \mbox{IMF--$\sigma$} relation to derive an IMF slope consistent with observations. Following \citet{labarbera}, we translated the central velocity dispersion of each galaxy in our sample to its expected IMF slope, as shown in Fig.~\ref{fig:francesco}.

\begin{figure}
\begin{center}
\includegraphics[width=8.5cm]{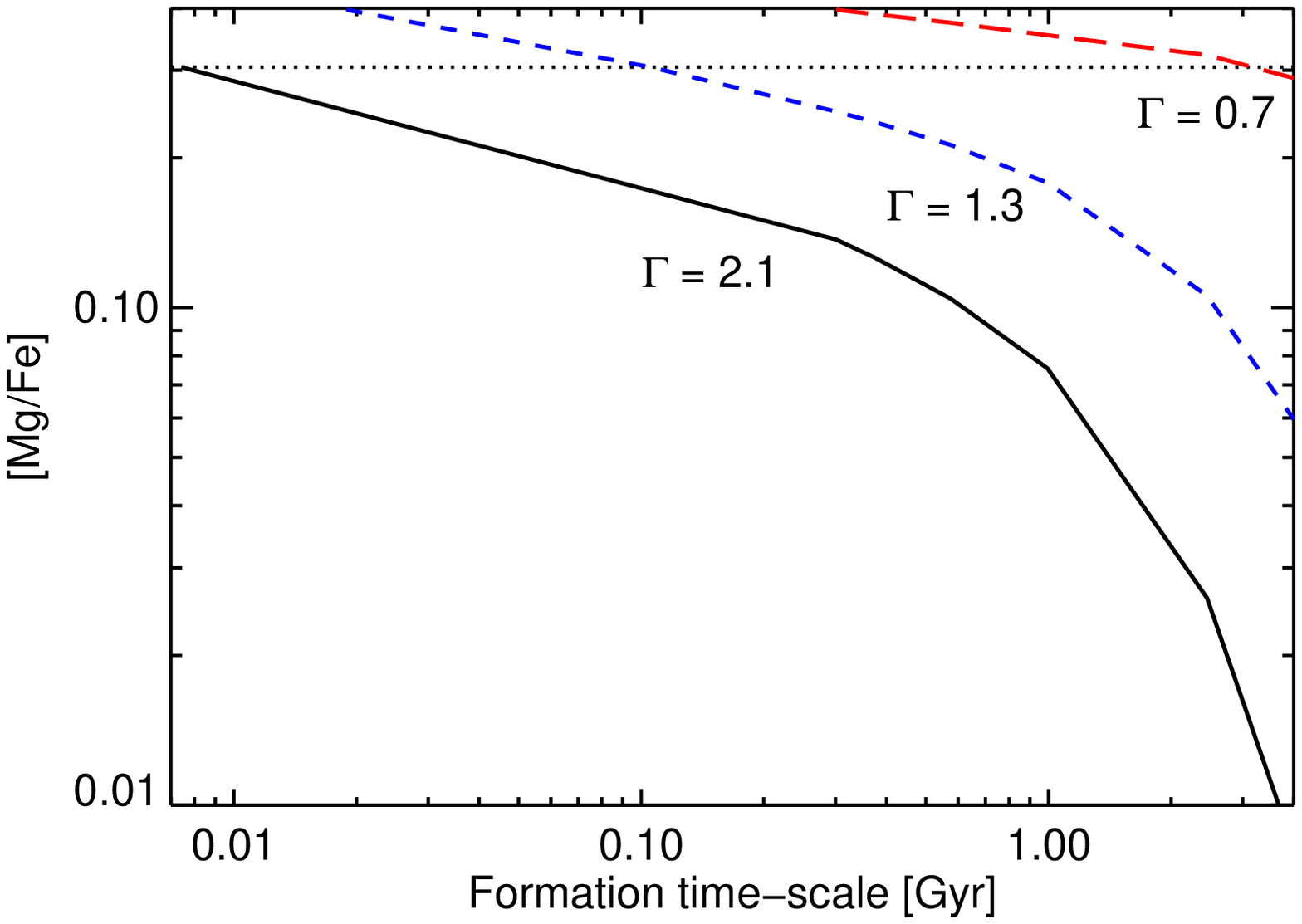}
\end{center}
\caption{The [Mg/Fe] proxy is represented as a function of the formation time-scale, for three different IMF slopes. Red an blue dotted line correspond to the clumpy collapse models described in \citet{Thomas99}, with the yields prescriptions of \citet{tnh96}. To match the observed IMF values in massive ETGs, we had to extrapolate these models to steeper IMFs. In this figure, we show the result of the extrapolation scheme for $\Gamma = 2.1$ (black solid), which is the typical IMF slope of very massive ($\sigma \gtrsim 300$\kms) ETGs. The horizontal dotted line ([Mg/Fe] = 0.3) correspond to the mean value for these massive ETGs. Therefore, the expected formation time-scale for galaxies with $\sigma \gtrsim 300$\kms is given by the intersection between the solid and dotted lines, corresponding in this case to $\tau = 0.007$ Gyr.}
\label{fig:alpha}
\end{figure}

The next step was to interpret the observed [Mg/Fe] ratios on the basis of these updated IMF values. Unfortunately, \citet{Thomas99} only explored models with flat ($\Gamma = 0.7$) and standard ($\Gamma = 1.3$) IMF slopes\footnote{This choice was motivated by the fact that a flat IMF slope was at the time a plausible explanation for the enhanced [Mg/Fe] ratios observed in massive ETGs.}, where $\Gamma$ corresponds to the slope of the IMF in $\log$-mass units. To overcome this problem, we extrapolated the models to steeper IMF values. Although a detailed chemical evolutionary model would be required for a more quantitative analysis, the qualitative behaviour of the [Mg/Fe] as a function of the IMF slope is reasonably well captured by the extrapolation. We expect the [Mg/Fe] to be a monotonic function of the IMF slope, since an IMF variation ultimately translates into a re-scaling of the number of massive stars that can pollute the inter stellar medium. In Fig.~\ref{fig:alpha} we show how the extrapolation scheme works for the most massive ETGs ($\Gamma \simeq 2$ or $\sigma \simeq 300$ \kms) in the sample.

\begin{figure}
\begin{center}
\includegraphics[width=8.5cm]{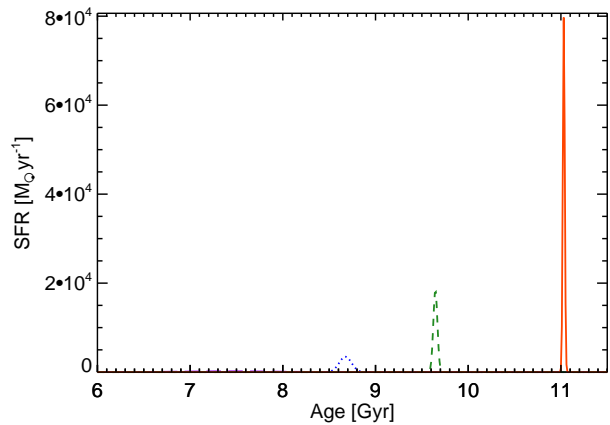}
\end{center}
\caption{Same plot as in Fig.~\ref{fig:thomas}, but now the width of each star formation history is calculated using the updated formation time-scales (Figs.~\ref{fig:francesco} and \ref{fig:alpha}). For the most massive galaxies in the sample ($\sigma \gtrsim 300$\kms), the updated $\tau$ value would be extremely short, certainly shorter than the recycling time needed to incorporate the processed material into the new generations of stars. Moreover, the expected star formation rate for these very massive ETGs would be orders of magnitudes (SFR$\sim10^5$\msun yr$^{-1}$) larger than the most extreme cases observed at high-redshift.}
\label{fig:thomod}
\end{figure}

Notice that a high [Mg/Fe] value can be reached either in a fast formation event or assuming a flat IMF. Since observations suggest that the IMF in massive ETGs is not flat, yet steeper than in the Milky Way, these updated formation time-scales are shorter than in \citet{thomas05}. Therefore, the more massive an ETG is, the steeper the IMF, the shorter the time-scale is compared to the standard view. Fig.~\ref{fig:thomod} shows how the star formation histories of ETGs with different masses would look like if [Mg/Fe] and IMF measurements are both taken into account.

The comparison between Fig.~\ref{fig:thomas} and Fig.~\ref{fig:thomod} reveals the strong influence of the IMF on the chemical evolution of ETGs. Whereas the star formation histories of low-mass ETGs, well modelled with a universal IMF, remain unaltered, massive ETGs require now extremely short and intense star-formation episodes. In particular, most massive galaxies in our sample ($\sigma \gtrsim 300$ \kms) have a typical IMF slope of $\Gamma = 2.1$ and a mean [Mg/Fe] value of 0.3 dex (Fig.~\ref{fig:alpha}). This would lead to formation episodes as fast as $\tau \approx 0.007$ Gyr, which implies star formation rates of $SFR \sim \mathrm{M_\star}/\tau = 1.1 \times 10^{5} \msun\,yr^{-1}$. Cooling and recycling all the ejected material in only $\sim 7$ Myr seems unreasonable \citep{Thomas99}. Moreover, the predicted star-formation rate would be orders of magnitude larger than observed in the most extreme submillimetre galaxies \citep{Barger}. Despite of the dust attenuation, objects forming $10^{5} \msun\,yr^{-1}$ should be easily detected at all redshift.

\section{Discussion}\label{sec:disc}

We have shown how the steep IMF slope and the enhanced [Mg/Fe] ratio observed in massive ETGs would imply unrealistically short formation time-scales. Thus, the use of 
[Mg/Fe] as a clock for galaxy formation warrants caution. 

It is worth noting that we have based our analysis only on the \citet{tnh96} yields prescriptions. However, if the \citet{ww95} models are used instead, the required formation time-scales would be even shorter \citep{Thomas99}. Fortunately, in the most recent yields models, there is a good agreement about the Mg production \citep{Matteucci}, and therefore it is unlikely that the IMF-[Mg/Fe] discrepancy is due to model uncertainties.

While the fact that massive ETGs have super-solar [Mg/Fe] ratios is well established and unanimously accepted, whether the IMF in these objects is universal or not is still under debate. There is a remarkable agreement between dynamical and stellar population IMF determinations, with almost completely independent approaches reporting similar trends with galaxy mass \citep{Treu,cappellari,ferreras13}. However, this agreement weakens when the comparison is made on a galaxy-by-galaxy basis \citep{Smith}. Only recently, stellar population and dynamical IMF analysis have been made over the same dataset, finding consistent results when both observables are taken into account \citep{lme,lb15}.

If yields models, IMF determinations and [Mg/Fe] measurements are in fact correct, two main scenarios have been proposed to reconcile the [Mg/Fe] and IMF observations:
\begin{list}{-}{}
  \item {\it Non-canonical IMF shape}. Nearby massive ETGs are old systems, and therefore, they are populated only by stars with masses below 1\msun. Thus, IMF studies are actually probing the IMF low-mass end. In contrast, the [Mg/Fe] ratio is settled by massive stars, being sensitive only to the high-mass end of the IMF. Then, in principle, it should be possible to construct an IMF with an enhanced (or standard) fraction of massive stars to match the observed [Mg/Fe] ratios, but also with an enhanced fraction of low-mass stars to reproduce the IMF observations. However, the large fraction of pristine gas locked in form of low-mass stars would prevent massive ETGs to reach their observed high-metallicities \citep{Ferreras15}. Moreover, under this parametrization, the low-mass IMF slope (below $\sim 0.5$M\sun) required to match the observed line-strength indices might lead to extremely high mass-to-light ratios, inconsistent with dynamical and lensing studies \citep{ferreras13, Ferreras15}.
  
  \item {\it Time-varying IMF}. Since IMF is a local property of ETGs \citep{mn14}, it is reasonable to argue that the IMF could evolve as subsequent generations of stars form in a galaxy, effectively altering its chemical properties, density and/or local pressure. In this regard, it has been suggested that a time-dependent IMF could explain at the same time the high metallicities, enhanced [Mg/Fe] ratios and steep IMFs measured in massive ETGs \citep{vazdekis:97,weidner:13,Ferreras15}. According to this scenario, the first generations of stars born within massive ETGs had an enhanced fraction of massive stars (i.e., a flat IMF). Then, a transition occurred, and massive ETGs continued to form stars but following a steep IMF. Thus, in the local Universe, the spectra of massive ETGs would be dominated by these latter populations, but with the chemical imprint of the first generations of stars. Although the mechanism responsible for this hypothetical IMF transition remains unknown, SFR- \citep{Gunawardhana} and metallicity-driven IMF variations \citep{mn15} have been suggested as plausible explanations. Note that a similar time-varying IMF have been implemented ad hoc in recent semi-analytical models to reconcile the evolution of stellar mass and luminosity functions \citep{Lacey}.
      
\end{list}

These two scenarios would lead to old, metal-rich stellar populations, with super-solar [Mg/Fe] ratios and an enhanced fraction of low-mass stars, as observed in nearby massive ETGs. However, the use of dynamical mass estimators could potentially break this observational degeneracy, since the amount of mass in form of stellar remnants, and therefore the mass-to-light ratio, might differ in both cases \citep[but see \S2.2 in][]{weidner:13}. In practice, systematic errors due to unconstrained dark matter halo profiles \citep{cappellari} and uncertainties in the IMF shape (Lyubenova et al. {\it in prep.}) prevent us from favouring one explanation over the other. Hence, to understand the chemical evolution of massive ETGs, it is necessary to observe them at high redshift while forming. 

Is it safe then to interpret the [Mg/Fe] ratio as an indicator of the formation time-scale of a galaxy? Surprisingly, [Mg/Fe]-based time scales are in great agreement with time-scales obtained through the analysis of star formation histories \citep{dlr,McDermid}, although in principle, they are completely independent. However, it is important to notice that it is not a well defined comparison. A star formation history consists on the summation of different single stellar populations (SSP), each of them with a certain age, metallicity and [Mg/Fe] ratio. Thus, it is always possible to produce narrow (fast) and extended (slow) star formation histories, but sharing the same luminosity weighted [Mg/Fe] value.

\section{Conclusions}

Galactic archaeology has provided deep insights on the formation and evolution of galaxies, thanks to the detailed study of nearby objects. However, without a proper knowledge of the chemical evolution of galaxies, a naive interpretation of $z\sim0$ observations could result into misleading conclusions. In particular, we have shown in this work that a joint consideration of the [Mg/Fe] ratio and IMF slope observed in nearby massive ETGs would lead to unrealistic formation time-scales and star formation rates. Therefore, one has to be careful while interpreting the [Mg/Fe] ratio purely as a tracer of the formation time-scale of a galaxy. 

Two main lines should be followed to relieve the tension between [Mg/Fe] ratios and formation time-scales. On the one hand, the non-universality of the IMF in massive ETGs must be further investigated. In particular, the discrepancies between dynamical and stellar population studies have to be understood before leaving behind the assumption of a universal IMF. Ultimately, it would be ideal to directly compare resolved and unresolved IMF determinations. Unfortunately, not even the upcoming James Webb Space Telescope will be able to resolve the central regions of massive enough ETGs, where the IMF is thought to depart from its Milky-Way shape \citep{mn14}. On the other hand, it is necessary to study {\it in situ} the formation of present-day massive ETGs. Thus, high-redshift observations, and in particular, a direct probe of the IMF during the formation epoch of massive galaxies will clarify the debate between a non-canonical or a time-varying IMF, as the explanation for the observed properties of nearby ETGs. The use of flexible stellar population synthesis models with a proper treatment of the abundance pattern \citep{cvd12,alpha}, combined with deep enough infrared data \citep{mosdef} will then be crucial over the next years.

\section*{Acknowledgements}

I would like to thank Prof. Jean P. Brodie for encouraging me to publish this work, and also to Alexandre Vazdekis, for making it possible. Also, I would like to thank Marja K. Seidel for the careful reading of the manuscript. Finally, I acknowledge the comments and suggestions from the referee, Dr Daniel Thomas, which I think greatly improved the discussion presented in this letter. I acknowledge support from the Spanish Government grant AYA2013-48226-C3-1-P.




\bibliographystyle{mnras}
\bibliography{alpha} 

\begin{thebibliography}{}
\makeatletter
\relax
\def\mn@urlcharsother{\let\do\@makeother \do\$\do\&\do\#\do\^\do\_\do\%\do\~}
\def\mn@doi{\begingroup\mn@urlcharsother \@ifnextchar [ {\mn@doi@}
  {\mn@doi@[]}}
\def\mn@doi@[#1]#2{\def\@tempa{#1}\ifx\@tempa\@empty \href
  {http://dx.doi.org/#2} {doi:#2}\else \href {http://dx.doi.org/#2} {#1}\fi
  \endgroup}
\def\mn@eprint#1#2{\mn@eprint@#1:#2::\@nil}
\def\mn@eprint@arXiv#1{\href {http://arxiv.org/abs/#1} {{\tt arXiv:#1}}}
\def\mn@eprint@dblp#1{\href {http://dblp.uni-trier.de/rec/bibtex/#1.xml}
  {dblp:#1}}
\def\mn@eprint@#1:#2:#3:#4\@nil{\def\@tempa {#1}\def\@tempb {#2}\def\@tempc
  {#3}\ifx \@tempc \@empty \let \@tempc \@tempb \let \@tempb \@tempa \fi \ifx
  \@tempb \@empty \def\@tempb {arXiv}\fi \@ifundefined
  {mn@eprint@\@tempb}{\@tempb:\@tempc}{\expandafter \expandafter \csname
  mn@eprint@\@tempb\endcsname \expandafter{\@tempc}}}

\bibitem[\protect\citeauthoryear{{Barger} et~al.,}{{Barger}
  et~al.}{2014}]{Barger}
{Barger} A.~J.,  et~al., 2014, \mn@doi [\apj] {10.1088/0004-637X/784/1/9},
  \href {http://adsabs.harvard.edu/abs/2014ApJ...784....9B} {784, 9}

\bibitem[\protect\citeauthoryear{{Bastian}, {Covey}  \& {Meyer}}{{Bastian}
  et~al.}{2010}]{bastian}
{Bastian} N.,  {Covey} K.~R.,   {Meyer} M.~R.,  2010, \mn@doi [\araa]
  {10.1146/annurev-astro-082708-101642}, \href
  {http://adsabs.harvard.edu/abs/2010ARA%26A..48..339B} {48, 339}

\bibitem[\protect\citeauthoryear{{Bender}, {Burstein}  \& {Faber}}{{Bender}
  et~al.}{1993}]{Bender}
{Bender} R.,  {Burstein} D.,   {Faber} S.~M.,  1993, \mn@doi [\apj]
  {10.1086/172815}, \href {http://adsabs.harvard.edu/abs/1993ApJ...411..153B}
  {411, 153}

\bibitem[\protect\citeauthoryear{{Calura} \& {Menci}}{{Calura} \&
  {Menci}}{2009}]{Calura}
{Calura} F.,  {Menci} N.,  2009, \mn@doi [\mnras]
  {10.1111/j.1365-2966.2009.15440.x}, \href
  {http://adsabs.harvard.edu/abs/2009MNRAS.400.1347C} {400, 1347}

\bibitem[\protect\citeauthoryear{{Cappellari} et~al.,}{{Cappellari}
  et~al.}{2012}]{cappellari}
{Cappellari} M.,  et~al., 2012, \mn@doi [\nat] {10.1038/nature10972}, \href
  {http://adsabs.harvard.edu/abs/2012Natur.484..485C} {484, 485}

\bibitem[\protect\citeauthoryear{{Conroy} \& {van Dokkum}}{{Conroy} \& {van
  Dokkum}}{2012}]{cvd12}
{Conroy} C.,  {van Dokkum} P.~G.,  2012, \mn@doi [\apj]
  {10.1088/0004-637X/760/1/71}, \href
  {http://adsabs.harvard.edu/abs/2012ApJ...760...71C} {760, 71}

\bibitem[\protect\citeauthoryear{{Ferr{\'e}-Mateu}, {Vazdekis}  \& {de la
  Rosa}}{{Ferr{\'e}-Mateu} et~al.}{2013}]{anna13}
{Ferr{\'e}-Mateu} A.,  {Vazdekis} A.,   {de la Rosa} I.~G.,  2013, \mn@doi
  [\mnras] {10.1093/mnras/stt193}, \href
  {http://adsabs.harvard.edu/abs/2013MNRAS.431..440F} {431, 440}

\bibitem[\protect\citeauthoryear{{Ferreras} et~al.,}{{Ferreras}
  et~al.}{2013a}]{ferreras13}
{Ferreras} I.,  et~al., 2013a, preprint, \href
  {http://adsabs.harvard.edu/abs/2013arXiv1312.5317F} {} (\mn@eprint {arXiv}
  {1312.5317})

\bibitem[\protect\citeauthoryear{{Ferreras}, {La Barbera}, {de la Rosa},
  {Vazdekis}, {de Carvalho}, {Falc{\'o}n-Barroso}  \&
  {Ricciardelli}}{{Ferreras} et~al.}{2013b}]{ferreras}
{Ferreras} I.,  {La Barbera} F.,  {de la Rosa} I.~G.,  {Vazdekis} A.,  {de
  Carvalho} R.~R.,  {Falc{\'o}n-Barroso} J.,   {Ricciardelli} E.,  2013b,
  \mn@doi [\mnras] {10.1093/mnrasl/sls014}, \href
  {http://adsabs.harvard.edu/abs/2013MNRAS.429L..15F} {429, L15}

\bibitem[\protect\citeauthoryear{{Ferreras}, {Weidner}, {Vazdekis}  \& {La
  Barbera}}{{Ferreras} et~al.}{2015}]{Ferreras15}
{Ferreras} I.,  {Weidner} C.,  {Vazdekis} A.,   {La Barbera} F.,  2015,
  preprint, \href {http://adsabs.harvard.edu/abs/2015arXiv150101636F} {}
  (\mn@eprint {arXiv} {1501.01636})

\bibitem[\protect\citeauthoryear{{Gunawardhana} et~al.,}{{Gunawardhana}
  et~al.}{2011}]{Gunawardhana}
{Gunawardhana} M.~L.~P.,  et~al., 2011, \mn@doi [\mnras]
  {10.1111/j.1365-2966.2011.18800.x}, \href
  {http://adsabs.harvard.edu/abs/2011MNRAS.415.1647G} {415, 1647}

\bibitem[\protect\citeauthoryear{{Kriek} et~al.,}{{Kriek}
  et~al.}{2014}]{mosdef}
{Kriek} M.,  et~al., 2014, preprint, \href
  {http://adsabs.harvard.edu/abs/2014arXiv1412.1835K} {} (\mn@eprint {arXiv}
  {1412.1835})

\bibitem[\protect\citeauthoryear{{Kroupa}}{{Kroupa}}{2002}]{kroupa}
{Kroupa} P.,  2002, \mn@doi [Science] {10.1126/science.1067524}, \href
  {http://adsabs.harvard.edu/abs/2002Sci...295...82K} {295, 82}

\bibitem[\protect\citeauthoryear{{La Barbera}, {Ferreras}, {Vazdekis}, {de la
  Rosa}, {de Carvalho}, {Trevisan}, {Falc{\'o}n-Barroso}  \&
  {Ricciardelli}}{{La Barbera} et~al.}{2013}]{labarbera}
{La Barbera} F.,  {Ferreras} I.,  {Vazdekis} A.,  {de la Rosa} I.~G.,  {de
  Carvalho} R.~R.,  {Trevisan} M.,  {Falc{\'o}n-Barroso} J.,   {Ricciardelli}
  E.,  2013, \mn@doi [\mnras] {10.1093/mnras/stt943}, \href
  {http://adsabs.harvard.edu/abs/2013MNRAS.433.3017L} {433, 3017}

\bibitem[\protect\citeauthoryear{{La Barbera}, {Vazdekis}, {Ferreras},
  {Pasquali}, {Cappellari}, {Martin-Navarro}, {Scoenebeck}  \&
  {Falcon-Barroso}}{{La Barbera} et~al.}{2015}]{lb15}
{La Barbera} F.,  {Vazdekis} A.,  {Ferreras} I.,  {Pasquali} A.,  {Cappellari}
  M.,  {Martin-Navarro} I.,  {Scoenebeck} F.,   {Falcon-Barroso} J.,  2015,
  preprint, \href {http://adsabs.harvard.edu/abs/2015arXiv150908250L} {}
  (\mn@eprint {arXiv} {1509.08250})

\bibitem[\protect\citeauthoryear{{Lacey} et~al.,}{{Lacey} et~al.}{2015}]{Lacey}
{Lacey} C.~G.,  et~al., 2015, preprint, \href
  {http://adsabs.harvard.edu/abs/2015arXiv150908473L} {} (\mn@eprint {arXiv}
  {1509.08473})

\bibitem[\protect\citeauthoryear{{Mart{\'{\i}}n-Navarro}, {Barbera},
  {Vazdekis}, {Falc{\'o}n-Barroso}  \& {Ferreras}}{{Mart{\'{\i}}n-Navarro}
  et~al.}{2015a}]{mn14}
{Mart{\'{\i}}n-Navarro} I.,  {Barbera} F.~L.,  {Vazdekis} A.,
  {Falc{\'o}n-Barroso} J.,   {Ferreras} I.,  2015a, \mn@doi [\mnras]
  {10.1093/mnras/stu2480}, \href
  {http://adsabs.harvard.edu/abs/2015MNRAS.447.1033M} {447, 1033}

\bibitem[\protect\citeauthoryear{{Mart{\'{\i}}n-Navarro}
  et~al.,}{{Mart{\'{\i}}n-Navarro} et~al.}{2015b}]{mn15}
{Mart{\'{\i}}n-Navarro} I.,  et~al., 2015b, \mn@doi [\apjl]
  {10.1088/2041-8205/806/2/L31}, \href
  {http://adsabs.harvard.edu/abs/2015ApJ...806L..31M} {806, L31}

\bibitem[\protect\citeauthoryear{{Matteucci}}{{Matteucci}}{2012}]{Matteucci}
{Matteucci} F.,  2012, {Chemical Evolution of Galaxies},
  \mn@doi{10.1007/978-3-642-22491-1.
}

\bibitem[\protect\citeauthoryear{{McDermid} et~al.,}{{McDermid}
  et~al.}{2015}]{McDermid}
{McDermid} R.~M.,  et~al., 2015, \mn@doi [\mnras] {10.1093/mnras/stv105}, \href
  {http://adsabs.harvard.edu/abs/2015MNRAS.448.3484M} {448, 3484}

\bibitem[\protect\citeauthoryear{{Peletier}}{{Peletier}}{1989}]{Peletier}
{Peletier} R.~F.,  1989, PhD thesis, , University of Groningen, The
  Netherlands, (1989)

\bibitem[\protect\citeauthoryear{{Schechter}, {Pooley}, {Blackburne}  \&
  {Wambsganss}}{{Schechter} et~al.}{2014}]{paul}
{Schechter} P.~L.,  {Pooley} D.,  {Blackburne} J.~A.,   {Wambsganss} J.,  2014,
  \mn@doi [\apj] {10.1088/0004-637X/793/2/96}, \href
  {http://adsabs.harvard.edu/abs/2014ApJ...793...96S} {793, 96}

\bibitem[\protect\citeauthoryear{{Smith}}{{Smith}}{2014}]{Smith}
{Smith} R.~J.,  2014, preprint, \href
  {http://adsabs.harvard.edu/abs/2014arXiv1403.6114S} {} (\mn@eprint {arXiv}
  {1403.6114})

\bibitem[\protect\citeauthoryear{{Spiniello}, {Trager}, {Koopmans}  \&
  {Conroy}}{{Spiniello} et~al.}{2014}]{Spiniello2013}
{Spiniello} C.,  {Trager} S.,  {Koopmans} L.~V.~E.,   {Conroy} C.,  2014,
  \mn@doi [\mnras] {10.1093/mnras/stt2282}, \href
  {http://adsabs.harvard.edu/abs/2014MNRAS.438.1483S} {438, 1483}

\bibitem[\protect\citeauthoryear{{Spiniello}, {Barnab{\`e}}, {Koopmans}  \&
  {Trager}}{{Spiniello} et~al.}{2015}]{lme}
{Spiniello} C.,  {Barnab{\`e}} M.,  {Koopmans} L.~V.~E.,   {Trager} S.~C.,
  2015, \mn@doi [\mnras] {10.1093/mnrasl/slv079}, \href
  {http://adsabs.harvard.edu/abs/2015MNRAS.452L..21S} {452, L21}

\bibitem[\protect\citeauthoryear{{Thielemann}, {Nomoto}  \&
  {Hashimoto}}{{Thielemann} et~al.}{1996}]{tnh96}
{Thielemann} F.-K.,  {Nomoto} K.,   {Hashimoto} M.-A.,  1996, \mn@doi [\apj]
  {10.1086/176980}, \href {http://adsabs.harvard.edu/abs/1996ApJ...460..408T}
  {460, 408}

\bibitem[\protect\citeauthoryear{{Thomas}, {Greggio}  \& {Bender}}{{Thomas}
  et~al.}{1999}]{Thomas99}
{Thomas} D.,  {Greggio} L.,   {Bender} R.,  1999, \mn@doi [\mnras]
  {10.1046/j.1365-8711.1999.02138.x}, \href
  {http://adsabs.harvard.edu/abs/1999MNRAS.302..537T} {302, 537}

\bibitem[\protect\citeauthoryear{{Thomas}, {Maraston}, {Bender}  \& {Mendes de
  Oliveira}}{{Thomas} et~al.}{2005}]{thomas05}
{Thomas} D.,  {Maraston} C.,  {Bender} R.,   {Mendes de Oliveira} C.,  2005,
  \mn@doi [\apj] {10.1086/426932}, \href
  {http://adsabs.harvard.edu/abs/2005ApJ...621..673T} {621, 673}

\bibitem[\protect\citeauthoryear{{Trager}, {Faber}, {Worthey}  \&
  {Gonz{\'a}lez}}{{Trager} et~al.}{2000}]{Trager00}
{Trager} S.~C.,  {Faber} S.~M.,  {Worthey} G.,   {Gonz{\'a}lez} J.~J.,  2000,
  \mn@doi [\aj] {10.1086/301299}, \href
  {http://adsabs.harvard.edu/abs/2000AJ....119.1645T} {119, 1645}

\bibitem[\protect\citeauthoryear{{Treu}, {Auger}, {Koopmans}, {Gavazzi},
  {Marshall}  \& {Bolton}}{{Treu} et~al.}{2010}]{Treu}
{Treu} T.,  {Auger} M.~W.,  {Koopmans} L.~V.~E.,  {Gavazzi} R.,  {Marshall}
  P.~J.,   {Bolton} A.~S.,  2010, \mn@doi [\apj]
  {10.1088/0004-637X/709/2/1195}, \href
  {http://adsabs.harvard.edu/abs/2010ApJ...709.1195T} {709, 1195}

\bibitem[\protect\citeauthoryear{{Vazdekis}, {Peletier}, {Beckman}  \&
  {Casuso}}{{Vazdekis} et~al.}{1997}]{vazdekis:97}
{Vazdekis} A.,  {Peletier} R.~F.,  {Beckman} J.~E.,   {Casuso} E.,  1997,
  \mn@doi [\apjs] {10.1086/313008}, \href
  {http://adsabs.harvard.edu/abs/1997ApJS..111..203V} {111, 203}

\bibitem[\protect\citeauthoryear{{Vazdekis} et~al.,}{{Vazdekis}
  et~al.}{2015}]{alpha}
{Vazdekis} A.,  et~al., 2015, \mn@doi [\mnras] {10.1093/mnras/stv151}, 449,
  1177

\bibitem[\protect\citeauthoryear{{Weidner}, {Ferreras}, {Vazdekis}  \& {La
  Barbera}}{{Weidner} et~al.}{2013}]{weidner:13}
{Weidner} C.,  {Ferreras} I.,  {Vazdekis} A.,   {La Barbera} F.,  2013, \mn@doi
  [\mnras] {10.1093/mnras/stt1445}, \href
  {http://adsabs.harvard.edu/abs/2013MNRAS.435.2274W} {435, 2274}

\bibitem[\protect\citeauthoryear{{Woosley} \& {Weaver}}{{Woosley} \&
  {Weaver}}{1995}]{ww95}
{Woosley} S.~E.,  {Weaver} T.~A.,  1995, \mn@doi [\apjs] {10.1086/192237},
  \href {http://adsabs.harvard.edu/abs/1995ApJS..101..181W} {101, 181}

\bibitem[\protect\citeauthoryear{{Worthey} \& {Collobert}}{{Worthey} \&
  {Collobert}}{2003}]{Worthey00}
{Worthey} G.,  {Collobert} M.,  2003, \mn@doi [\apj] {10.1086/367607}, \href
  {http://adsabs.harvard.edu/abs/2003ApJ...586...17W} {586, 17}

\bibitem[\protect\citeauthoryear{{Worthey}, {Faber}  \& {Gonzalez}}{{Worthey}
  et~al.}{1992}]{Worthey92}
{Worthey} G.,  {Faber} S.~M.,   {Gonzalez} J.~J.,  1992, \mn@doi [\apj]
  {10.1086/171836}, \href {http://adsabs.harvard.edu/abs/1992ApJ...398...69W}
  {398, 69}

\bibitem[\protect\citeauthoryear{{de La Rosa}, {La Barbera}, {Ferreras}  \& {de
  Carvalho}}{{de La Rosa} et~al.}{2011}]{dlr}
{de La Rosa} I.~G.,  {La Barbera} F.,  {Ferreras} I.,   {de Carvalho} R.~R.,
  2011, \mn@doi [\mnras] {10.1111/j.1745-3933.2011.01146.x}, \href
  {http://adsabs.harvard.edu/abs/2011MNRAS.418L..74D} {418, L74}

\bibitem[\protect\citeauthoryear{{van Dokkum} \& {Conroy}}{{van Dokkum} \&
  {Conroy}}{2010}]{vandokkum}
{van Dokkum} P.~G.,  {Conroy} C.,  2010, \mn@doi [\nat] {10.1038/nature09578},
  \href {http://adsabs.harvard.edu/abs/2010Natur.468..940V} {468, 940}

\makeatother
\end{thebibliography}


\bsp	
\label{lastpage}
\end{document}